\documentstyle[seceq,letter,epsf,wrapft]{ptptex}

\title{
Response to Invasion by Antigen and Effects of Threshold
in an Immune Network Dynamical System Model
 with a Small Number of
Degrees of Freedom
}

\author{
Satoko {\sc Itaya} and Tatsuya {\sc Uezu}}

\inst{
Graduate School of Human Culture, \\
Nara Women's University, Nara 630-8506, Japan}

\recdate{
}

\abst{
We study a dynamical system model 
of an idiotypic immune network with a small number of
degrees of freedom, mainly focusing on the effect of a threshold
above which antibodies can recognise antibodies.
The response of the system to
invasions by antigens is investigated
 in the both models with and without the threshold
 and it turns out that the system changes in a desirable
direction for moderate magnitude of perturbation.
Also, the propagation of disturbance by an antigen is investigated 
in the system of one-dimensionally connected basic units
taking the closed 3-clone system as a unit, and it is clarified
that the threshold of the system has effects to enhance the stability of
 the network and to localise the immune response.
}

\begin{document}
\maketitle
\noindent
{\bf Introduction}\par
\normalsize
  It is experimentally well-known that an immune system is activated
by interacting with itself  to prepare for unknown 
antigens\cite{exp1,exp2}.  Taking this fact into account,
an immune network model was introduced by N. K. Jerne\cite{Jerne 74}.
Later, a dynamical system model of the immune network was introduced by
 F. J. Varela et. al.\cite{Varela 88}.
In this model, the basic elements are antibodies and 
B-cells which produce antibodies.
 The effects of helper and suppressor T-cells are taken into account
by introducing functions which represent 
 interactions between antibodies, and between B-cells and antibodies.
This has been called  the ``second generation immune network model''
\cite{Stewart Varela 89}.
In this model, some important mechanisms such as recognition, memory, and 
tolerance have been studied
\cite{Stewart Varela 89,Varela 91,Calenbuhr et.al. 95,Calenbuhr et.al. 96}.

In the previous paper\cite{itaya 99}, we studied
the original Varela model and also the modified model.
Mainly, we have investigated the model with a threshold 
above which antibodies can recognise antibodies  and have reported
that the system has chaotic states,  and
also a peculiar type of limit cycle, which we called differentiating state.
Further, as for the response to the invasion by antigens, 
we have found that when the system is in a differentiating state,
its response to a specific type of antigen is sensitive. 
That is, differentiating states are considered to represent a kind of 
short term memory of the invasion by  antigens.
Further, we have found that the response time in a chaotic state
  takes an intermediate value compared to differentiating states. \par
An important issue which has not yet been addressed is the
question of how central  the use of thresholds is to the results found.
This is investigated in the first part of the present paper.
In the second part of the paper we go on to study  the localisation
of immune responses. 
It has been pointed out by De Boer et.al. that
 a fault of the network view of immune systems is that 
when only one clone interacts with one antigen,
immune responses spread over the whole system 
and the activation by the invasion by antigens is
not localised in the model by Varela et.al. 
This behaviour seems inappropriate 
because it suggests that large networks, such as might be found
in the real immune system, can be unstable since small local  
 perturbations can generate global effects by spreading over 
 the whole system.  Thus, it is very interesting to find a mechanism 
to remove this fault and to localise
the immune responses\cite{Boer 96}. Thresholds may be one of such candidate.
Therefore, we study the effect of thresholds on
the response to the invasion by antigens,  by investigating
the propagation of disturbances in one-dimensionally-connected basic units.\\

\noindent
{\bf Model}\par
The evolution in time for the concentrations of an antibody, $f_i$,
 and of a B-cell, $b_i$, with the idiotype $i(i=1, 2, \cdots, N)$
are given by the following equations.
\begin{eqnarray}
df_{i}/dt &=&
-K_{1}\sigma_{i}f_{i}-K_{2}f_{i}+K_{3}Mat
\left(\sigma_{i}\right)b_{i},
\label{eqn:F}\\
db_{i}/dt &=&
-K_{4}b_{i}+K_{5}Prol\left(\sigma_{i}\right)b_{i}+K_{6}.
\label{eqn:B}
\end{eqnarray}
\begin{wrapfigure}{l}{\halftext}
\epsfxsize=4cm
\centerline{\epsfbox{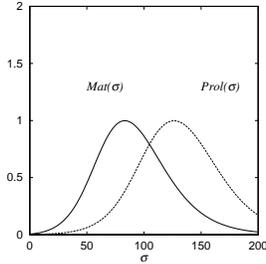}}
\caption{Profiles of $Mat(\sigma)$ and $Prol(\sigma)$ adopted in this paper}
\label{fig:MatProl}
\end{wrapfigure}
See Ref. 9) for detailed definitions and explanations.
Here, $\sigma_{i}$ is the sensitivity 
of the network to the $i$th idiotype, defined as
\begin{eqnarray}
\sigma_{i}=\mathop{\sum}_{j=1}^{N}m_{ij}^0f_{j}.
\end{eqnarray}
The interaction between two different idiotypes $i$ and $j$ are
represented by the connectivity $m_{ij}^{0}$.
We set $m_{ij}^{0}>0$ if there is an affinity between $i$ and $j$, and
$m_{ij}^{0}=0$ if there is not.  For simplicity, we assume
$m_{ij}^0=1 - \delta_{i,j}$, where $\delta_{i,j}$ is the Kronecker's delta. 
$Mat\left(\sigma_{i}\right)$
and $Prol\left(\sigma_{i}\right)$ are the interaction functions
between antibodies, and antibodies and B-cells, respectively.  
In general,  $Mat\left(\sigma_{i}\right)$ and
$Prol\left(\sigma_{i}\right)$ are assumed to be unimodal functions and 
 $Prol\left(\sigma_{i}\right)$  to be  shifted to the right with respect
to $Mat\left(\sigma_{i}\right)$ in order 
to take the roles of T-cells into account. 
The functions we adopt here are the followings.
See Fig.\ref{fig:MatProl}. 
\begin{eqnarray}
\hspace*{-10mm}
Mat \left( \sigma_{i}\right) &=& U_{1}
\left[\tanh \left\{ U_{2} \left(\sigma_{i}-T_{lm}\right)\right\}
 -\tanh \left\{ U_{3}\left(\sigma_{i}-T_{um}\right)\right\}\right],
\label{eqn:4}\\
\hspace*{-10mm}
Prol\left(\sigma_{i}\right)&=&U_{4} \left[\tanh\left\{U_{5} \left(
\sigma_{i} -T_{lp}\right)\right\}
 - \tanh\left\{U_{6} \left(\sigma_{i}-T_{up}
\right)\right\}\right].
\end{eqnarray}
\noindent
{\bf Model with Threshold}\par
First, let us study  a 3-clone closed chain system with a threshold 
$\kappa_{0}$ above which the $i$th antibody can recognise other
antibodies.  Each element of the connectivity matrix $M=\{ m_{ij} \}$ 
is defined as
\begin{equation}
m_{ij}(t) = m_{ij}^{0}\Theta(f_j(t)-\kappa_0),
\end{equation}
where $\Theta(x)$ is the Heaviside function (i.e.,$\Theta(x)=1$ for $x
\geq 0$ and 0 for $x < 0$).
In this system, there exist differentiating states\cite{itaya 99}.
We consider the response of the system in differentiating states
 to invasion by antigens.  
In these states there exist a  clone with a long period
and a clone with a short period.  We denote  the former by
L, and the latter by S, respectively.  
We treat two cases of invasion by antigens in the followings.\\
\underline{Case 1}\par
An antigen $A$ similar to the
antibody $f_{1}$ invades
 the system.  We assume in this case that 
antibodies $f_{2}$ and $f_{3}$ recognise the 
antigen as  the antibody $f_{1}$.  Thus,
in the sensitivities $\sigma_2$ and $\sigma_3$,
 $f_1$ is replaced by $f_{1}'=f_{1}+A$. 
Further, thresholds related to the antigen are also set to $\kappa _0$.
On the other hand, in general, the antibody $f_{1}$ cannot
recognise the antigen.
Then, by assuming that the antigens enter the system at a rate
$K_{7}$ per unit time, the differential equation for the antigen $A$ is
given by
\begin{wrapfigure}{r}{\halftext}
\epsfxsize=4cm
\centerline{\epsfbox{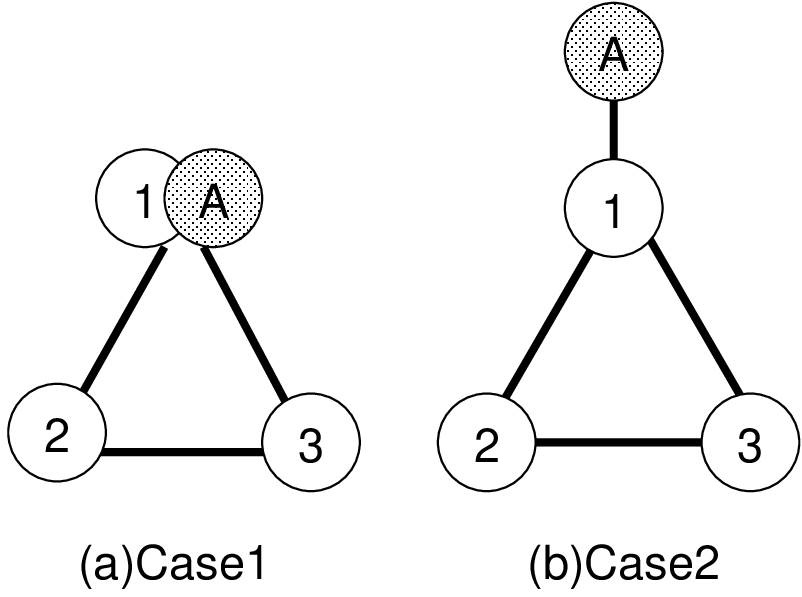}}
\label{fig:anti}
\caption{invasion of antigen}
\end{wrapfigure}
\begin{eqnarray}
\frac{dA}{dt}& =& -K_{1}\sigma_A(t) A+K_{7},\\
&& \sigma_A(t)=m_{12}(t)f_{2}(t)+m_{13}(t)f_{3}(t). \nonumber
\end{eqnarray}
\underline{Case 2}\par
An antigen $A$ interacts only with the antibody $f_{1}$.
Then, the antibodies $f_{1}$ recognise the antigen, but
antibodies $f_{2}$ and $f_{3}$ cannot recognise the antigen,
 in general.
Let us define $\kappa_{i,j}$ as the threshold above which 
$f_j$ recognises $f_i$.
Here $i$ and $j$ take integer values between 1 and 4 where
 $i=4$ denotes the antigen.
Then the equation for the antigen $A$ is
\begin{eqnarray}
\frac{dA}{dt}&=&-K_{1}\sigma_A(t)A+K_{7},\\
&&\sigma_A(t)=\Theta(f_1 - \kappa_{4, 1})f_1.\nonumber
\end{eqnarray}
The sensitivity $\sigma_{i}$ of the $i$th clone is modified as
\begin{equation}
\sigma_{i}=\sum_{j=1} ^3 m_{ij}^0 \Theta(f_{j}-\kappa _{i, j}) f_j
+ l_i \Theta(A - \kappa_{i, 4}) A,
\end{equation}
where $l_{i}$ is the strength of the interaction between $f_{i}$ and
$A$ and is set to $l_{i}= s_{A} \delta _{i,1}$ with $s_A=0.3$.
We assume $\kappa_{i,j}=\kappa_{j,i}= \kappa _0 =50$ for
any  $i, j=1, 2, 3$ and $\kappa_{i,4}=\kappa_{4,i}= \kappa _1 =10$ for
 $i=1, 2, 3$.\par
We have performed numerical calculations in these two cases
and have found the following results about the response of the system.
As is shown in Ref. 9), there is a differentiating state
with two L clones and one S,
which responds to the antigen much better 
than the other states.
In both cases 1 and 2, if the input rate $K_7$ of the antigen 
is neither very small nor very large, the system modifies itself
by evolving to that differentiating state in which 
the relaxation time of the antigen is shortest among differentiating states.
This phenomenon is regarded as a kind of short term memory of the
invasion by the antigen.  
However, in both cases, 
if $K_{7}$ is sufficiently large, the concentration of antibody which can
interact with the antigen converges to a fixed point, while 
the antigen concentration grows exponentially 
with the result that the system is destroyed by the antigens.
The dependence on the initial condition of this behaviour
is less sensitive than that on $K_7$. 
Further, we have found that  the network is destroyed by the antigen
more easily in the case 2 than in the case 1.\\

\noindent
{\bf Model without Threshold}\par
\begin{wrapfigure}{e}{\halftext}
\epsfxsize=6cm
\centerline{\epsfbox{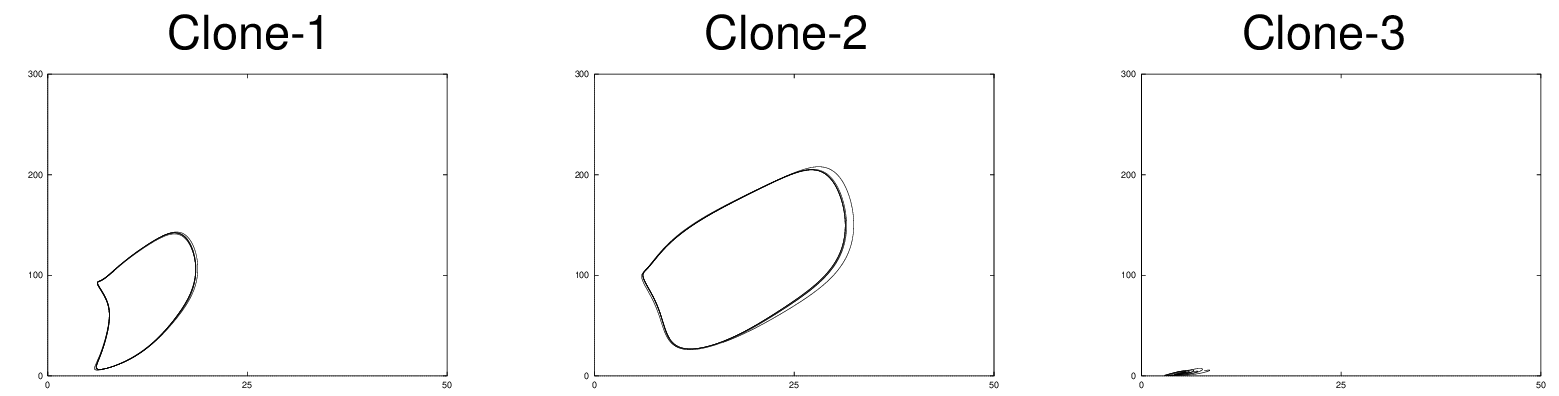}}
\caption{Limit Cycle 3}
\label{fig:limit3}
\end{wrapfigure}

\normalsize
Next, we study the response of the system without the threshold to invasion
by the antigen in  cases 1 and 2.
We introduce a strength parameter $s$ of the connectivity, 
and set the elements of the connectivity matrix as
\begin{eqnarray}
m_{ij}=s m_{ij} ^0=s( 1- \delta_{i,j}).
\label{eqn:MC}
\end{eqnarray}
As $s$ increases from 0, the system changes as
a fixed point $\rightarrow $ limit cycle 1
$\rightarrow $ chaos $\rightarrow $ limit cycle 2
$\rightarrow $ a fixed point.
If the state is either limit cycle 2 or chaotic, 
the whole network is activated.  On the other hand,
 in the limit cycle 1 state, a part of the network is activated, 
in the sense that one clone takes
negligibly small values and does not affect the other clones.
By fixing the initial value of the antigen and changing  $K_{7}$, 
we study the  responses of these three states.
In both Case 1 and Case 2,
for a small value of $K_{7}$, the state of the system 
does not change very much,
but for large $K_{7}$, 
 each clone converges to a fixed point and the network is destroyed.
The magnitude of the value of $K_{7}$ that causes the
breakdown of the system is of order $1$
in Case 1, and is $10 ^{-1}$ in Case 2.
In Case 1, for intermediate values of $K_{7}$,
regardless of initial states,
 the attractor changes to a new periodic state(limit cycle 3)
 (Fig.\ref{fig:limit3}).
This state is considered to be a better state against 
the invasion by the antigen 
because one of two clones which can interact with the antigen is
in the long-period state, L.
In Case 2,
when $K_{7}$ takes intermediate values,
the attractor changes to the  chaotic state
in which the clone 1 oscillates in relatively longer pulse states.
These behaviours occur for all initial states except for 
the state of the limit cycle 1 in which the clone 1 takes small values. 
Thus, as well as in Case 1,  in Case 2
 the system tends to take  better configurations 
to deal with the antigen.\\

Therefore,  whether there is the threshold or not,
it turns out that in both cases 1 and 2,
the system changes to a desirable
direction for moderate magnitudes of perturbation.
  Further, it turns out that the network
is destroyed by the antigen more easily in Case 2 than in Case 1.
The reason  seems to be that
 the effect of the antigen on the network is more direct
in Case 2  than in Case 1.\\

\noindent
{\bf 
Loosely Connected 3-clone Units}\par
Here, to study the effect of thresholds on
the response to the invasion by antigens,  we investigate 
one-dimensionally connected basic units, 
where each basic unit is composed of a 3-clone closed system.\\
\noindent
\underline{ A. Basic units with threshold}\par
\begin{figure}[htb]
\epsfxsize=12cm
\centerline{\epsfbox{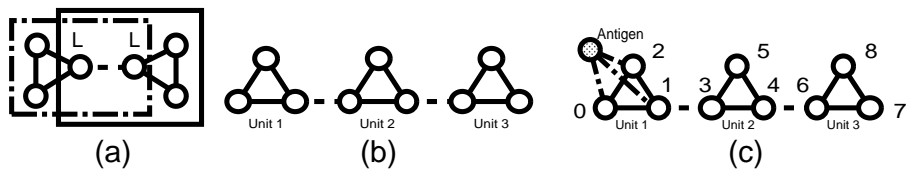}}
\caption{Connection of 3-clone units}
\label{fig:BASE_U}
\end{figure}

As an initial state of a basic unit, we take
a limit cycle in which the configuration of clones is (L, L, S).
When two basic units are connected, for each unit the network architecture
is similar to Case 2 of the invasion by the antigen.
See Fig.\ref{fig:BASE_U}(a). 
Thus, from the above result  in Case 2, if the interaction
between units is large, the combined system is destroyed when
the amplitude in either unit becomes large.
Hence, hereafter, we consider loosely connected systems.
\begin{wrapfigure}{e}{\halftext}
\epsfxsize=6cm
\centerline{\epsfbox{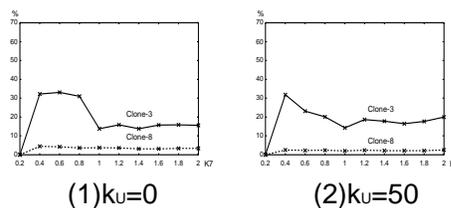}}
\caption{$K_{7}$ dependence of the magnitude of fluctuation in clone 3 and 8
relative to that in clone 0}
\label{fig:LAST_1}
\end{wrapfigure}
In the case of two units, the clone which 
is connected to the  other unit tends to stay in the long pulse state,
and the basin of attraction of this type of solution becomes large. 
It turns out that this is a characteristic feature 
for the loosely connected systems. 
Now, let us study an open chain of 
three basic units:  unit 1, unit 2 and unit 3.  
See Fig.\ref{fig:BASE_U}(b). 
 Let $s_U$ and $\kappa_U$ be the strength of the connection between units 
and that of the threshold between units, respectively.
To see the response behaviour quantitatively, 
we have calculated in limit cycle states 
the differences between the phases
of oscillations in the presence and the absence of the antigens.
We have also changed $\kappa _0$ (the threshold between clones) 
and $\kappa _U$.
As a result, we have observed that
the disturbance is reduced more in unit 3 than in unit 2.
However, significant difference  between in the cases
of $\kappa _U=0$ and  $\kappa _U>0$ has not been observed.
See Fig.\ref{fig:LAST_1}.  Further, we have found 
that the disturbance is not always reduced
as $\kappa _0$ is increased, and that
 the threshold between clones provokes complicated dynamical 
behaviours.  Therefore, to see the effect of the threshold more clearly,
we investigate the system composed of the basic units without thresholds.\\
\noindent
\underline{ B. Basic units without threshold}\par
\begin{wrapfigure}{e}{\halftext}
\epsfxsize=6cm
\centerline{\epsfbox{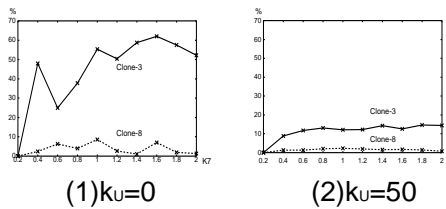}}
\caption{$K_{7}$ dependence of the magnitude of fluctuation in clone 3 and 8
relative to that in clone 0}
\label{fig:LAST}
\end{wrapfigure}

We set $\kappa_0=0$.
In this case, for $s=2.5$ and $s_U=0.05$, there exist limit cycle states
in both cases of $\kappa_U=0$ and 50.
As in case A,  setting $A_0=80, \kappa_1=10$ and $s_A=0.3$,
we have investigated the response behaviour of
these limit cycles quantitatively.  
We have observed that the disturbance again hardly spreads to unit 3.
 Further, we have found that 
the disturbance in unit 2 is reduced considerably for $\kappa_U=50$,
while it is still large for $\kappa_U=0$.  See Fig.\ref{fig:LAST}. \\

These results show that the thresholds 
have the effects of reducing the  magnitudes of  interactions
substantially, and of enhancing the independence of each unit.
\noindent
{\bf Discussion}\par
In this letter, we have analysed the model constructed from
the point of view
of N. K. Jerne that antibodies look upon other
antibodies as internal images of antigens. 
It has been pointed out as a big fault in his network theory
 that the stability of the network is not guaranteed. 
There have been several studies 
on this subject\cite{Varela 91,Calenbuhr et.al. 96,Varela Stewart 90}.
However, it seems that this problem  has not yet been fully
investigated.  The result on immune responses 
 obtained in the first part of this paper
implies that the system does not always change randomly
 but can change in an ordered way in some cases.  This 
 may be a refutation of the above criticism of the network theory. 
Since this result has been obtained in systems
with few degrees of freedom, it is necessary to study
 larger systems. This is now under investigation.\par
Now, let us discuss the effects of thresholds of concentrations.
The result in the second part of this paper shows that
the structural stability of the network is increased by the
existence of thresholds.
It has been pointed out that 
a cross-linking structure made by antibody molecules 
is very important in order that phagocytes can catch antigens 
in real immune systems\cite{TheCell}.  
It is natural to assume that
the formation of the three-dimensional cross-linking structure
by antibodies depends on the  concentrations of
antibodies and antigens. Thus, the introduction of 
the threshold of the concentrations
not only gives a desirable effect which makes the
network  more stable, but also 
can be interpreted as 
taking one of the mechanisms of 
real immune responses into account theoretically.
It would be very interesting to know whether the threshold provokes  the
 localisation of the immune response 
in larger systems.  
This is a future problem.\\

We are grateful to D. Broomhead and J. Huke 
for critical reading of the manuscript.

\end{document}